\documentclass[sigconf, nonacm]{acmart}

\usepackage{microtype}
\usepackage{booktabs}
\usepackage{amsmath}
\usepackage{float}
\usepackage{subcaption}
\usepackage{indentfirst}
\usepackage{tabularx}
\usepackage{threeparttable}
\usepackage{makecell}

\newtheorem{assumption}{Assumption}

\title{GitHub Copilot and Developer Productivity:\\ An Observational Dose-Response Analysis}

\author{Alex Heilman}
\affiliation{%
  \institution{Microsoft, Engineering Thrive}
  \city{Redmond}
  \state{WA}
  \country{USA}}
\email{alexheilman@microsoft.com}

\author{Alex Kyllo}
\affiliation{%
  \institution{Microsoft, Engineering Thrive}
  \city{Redmond}
  \state{WA}
  \country{USA}}
\email{jekyllo@microsoft.com}

\author{Emerson Murphy-Hill}
\affiliation{%
  \institution{Microsoft, CHISE}
  \city{Mountain View}
  \state{CA}
  \country{USA}}
\email{emerson.rex@microsoft.com}

\begin{document}

\begin{abstract}
Does GitHub Copilot (GHCP) make engineers more productive, or do the engineers who use it more differ from those who use it less? And even within a single engineer, are GHCP-heavy weeks just busy weeks in which more of everything gets done? We study these questions using 43 weeks of data from 16,223 software engineers across Microsoft's Cloud+AI organization. Engineer fixed effects address the first concern by comparing each engineer against themselves rather than against other engineers, eliminating time-invariant differences in skill, role, and team. Active coding time and browser time then enter a Poisson Pseudo-Maximum Likelihood model with two-way fixed effects to address the harder, within-engineer confound: that GHCP-heavy weeks coincide with high-effort weeks. This defines our estimand as an efficiency effect: more pull requests completed at equivalent levels of coding time. Engineers are estimated to complete 40.5\% more PRs in their highest GHCP usage weeks relative to their zero-usage weeks, holding measured development effort constant. The gradient is monotonic with diminishing returns at high intensity. Seven robustness and falsification tests target the remaining plausible alternative explanations (non-coding AI engagement, team-level shocks, within-week task reallocation, cross-week contamination, PR slicing into smaller units, shifts toward easier task types, and sensitivity to how the treatment is operationalized). Under an explicitly stated conditional-independence assumption, the within-engineer design estimates a tool-specific efficiency effect that is consistent with all seven robustness tests.
\end{abstract}



\maketitle
\pagestyle{plain}

\section{Introduction \& Motivation}
\label{sec:introduction}

Measuring the productivity impact of LLM coding assistants is deceptively difficult. A growing empirical literature examines how Copilot affects developer work, including controlled experiments by Peng et al.\ \cite{peng2023}, quasi-experiments on task allocation by Hoffmann et al.\ \cite{hoffmann2025}, and randomized field experiments by Cui et al.\ \cite{cui2025} reporting productivity gains. Extending these findings from initial adoption to the steady-state productivity of a ubiquitous tool requires a new methodology.

\textbf{Why not an A/B test?} The gold standard for causal inference is not feasible here. GHCP is already widely deployed, and removing an established tool from a random subset of engineers raises practical and ethical concerns. This would disrupt workflows, drive engineers to an alternative, and measure the cost of losing a tool rather than the benefit of having one. One could A/B test an incremental GHCP feature (e.g., a new model version), but that only measures the marginal improvement, not the overall productivity impact of GHCP.

\subsection{Two Layers of Confounding}

The motivating question hides two distinct confounds, and our design addresses them in two stages.

\textbf{Between-engineer selection.} The engineers who use GHCP more may differ systematically from those who use it less---more skilled, on different teams, working on PR-friendly task mixes, or simply more receptive to new tools. Any cross-sectional comparison conflates these differences with the tool's effect.

\textbf{Within-engineer, time-varying confounding.} Even for a fixed engineer, the weeks in which they use GHCP heavily may not be comparable to the weeks in which they use it lightly. A heads-down crunch week, a large feature landing, or a productive stretch could plausibly raise both GHCP usage and PR output for reasons unrelated to the tool's marginal contribution.

\subsection{Approach: Within-Engineer Dose-Response}

This analysis exploits within-engineer, week-to-week variation in GHCP usage intensity to address both layers. Engineer fixed effects fully absorb the time-invariant between-engineer selection problem. We compare each engineer against themselves, not against other engineers, so any time-invariant difference in ability, role, team, or task allocation is differenced out. Active coding time and browser time then address the harder within-engineer confound by holding measured development effort constant in the same week. Week fixed effects absorb shocks shared across the population.

\begin{assumption}[Conditional independence]\label{assump:identification}
Conditional on engineer fixed effects, week fixed effects, coding time, and browser time, an engineer's GHCP usage intensity is not systematically driven by the same unobserved factors that independently determine their PR output that week.
\end{assumption}

Like any non-experimental identification, this assumption cannot be tested directly. Section~\ref{sec:robustness} probes its plausibility through a seven-test falsification battery, and Section~\ref{sec:discussion} characterizes the threats it cannot rule out.

\newpage
\subsection{Causal Inference Context}

Pearl's ``Ladder of Causation'' distinguishes three levels of causal reasoning, each requiring progressively stronger evidence \cite{pearl2018}.

\begin{enumerate}
\item \textbf{Association:} observing that GHCP users complete more PRs.
\item \textbf{Intervention:} what would happen if we \emph{made} an engineer use GHCP more? Only a randomized experiment answers this directly.
\item \textbf{Counterfactual:} what \emph{would have} happened to this specific engineer in this specific week had they not used GHCP?
\end{enumerate}
A randomized A/B test would place us on rung~2. That is not feasible here for the reasons described above. We therefore operate on rung~1, but with design choices that narrow the gap toward rung~2:

\begin{itemize}
\item \textbf{Within-engineer comparison.} Engineer fixed effects eliminate all time-invariant confounds (ability, experience, team, role, coding style). We compare the \emph{same person} across weeks, not different people.
\item \textbf{Calendar fixed effects.} Week fixed effects absorb time-varying shocks that are common to the population (deadlines, holidays, reorgs).
\item \textbf{Effort controls as estimand choice.} Coding time and browser time enter the model as controls, which defines the estimand as an \emph{efficiency} effect: the association between GHCP usage and PR output, holding measured development effort constant. If GHCP reduces the time needed to complete a task, that channel is absorbed by the controls, making our estimates conservative (see Section~\ref{sec:estimand}).
\item \textbf{Falsification battery.} Sections~\ref{sec:placebo-treatment}--\ref{sec:breadth} systematically test whether the remaining time-varying confounds can explain the gradient.
\end{itemize}

\section{Methodology}
\label{sec:methodology}

\subsection{The Confounded Baseline}

Before introducing the specification, we establish the naive comparison as motivation. The simplest question one might ask is whether engineers who use GitHub Copilot more complete more PRs, and at the raw weekly average level the answer is yes. That answer is not informative, however, because engineers who use GHCP more also tend to spend more time coding in those weeks. Any positive cross-sectional association between usage and PR output is mechanically confounded by effort, and we cannot tell from such a comparison whether GHCP is contributing to productivity or whether we are merely observing a correlation with hours worked. This motivates the specification developed in the remainder of the section.

\subsection{Fixed-Rate Offset Model}
\label{sec:offset}

A natural starting point is to model productivity as a rate: completed PRs per active coding hour. We accomplish this by placing $\ln(\text{CodingTime})$ as an offset in a Poisson regression:

\begin{align*}
\ln E\!\left[\frac{\text{PRs}_{it}}{\text{CodingTime}_{it}}\right] = \beta_1 \cdot \text{Low}_{it} + \beta_2 \cdot \text{Mod}_{it} \\
 + \beta_3 \cdot \text{High}_{it} + \delta_t + \alpha_i
\end{align*}

This forces a proportionality assumption, in which doubling coding hours must double the PRs. Each $\beta$ then measures the percentage change in the PR rate at a given GHCP usage level relative to no usage, neatly isolating an efficiency effect \emph{if} the assumption holds.

\begin{figure}[H]
\begin{center}
\centerline{\includegraphics[width=\columnwidth]{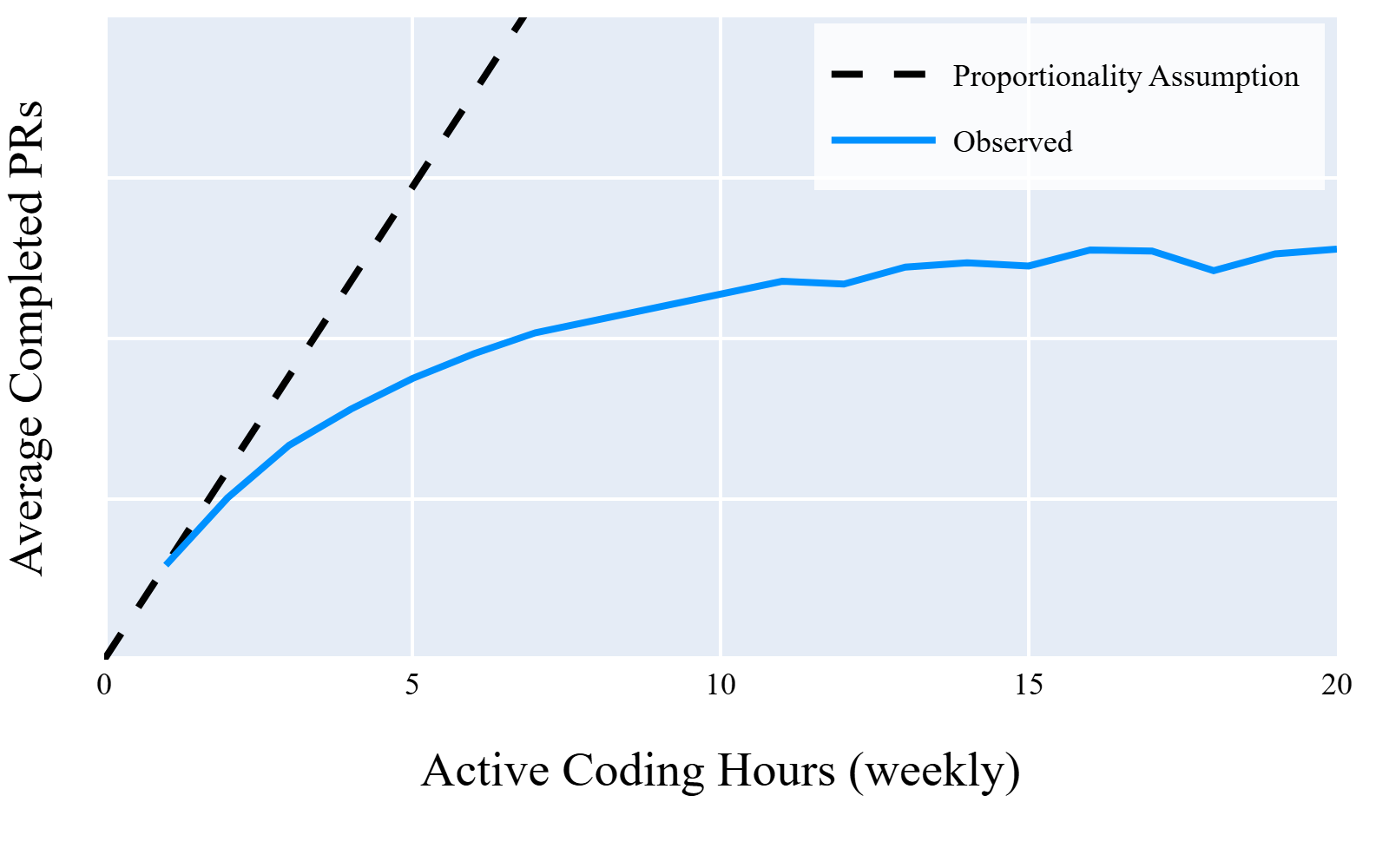}}
\caption{Average PRs vs.\ active coding hours, with proportional reference line.}
\label{fig:proportionality}
\end{center}
\end{figure}

The assumption is empirically false in our panel. Figure~\ref{fig:proportionality} plots average PRs against active coding hours alongside the proportional reference line the offset spec assumes. The observed curve falls below that line and flattens as hours rise, consistent with diminishing returns to effort. Because GHCP usage and coding hours are themselves correlated in the panel, any apparent GHCP coefficient from this specification absorbs the curvature in the hours--PRs relationship rather than isolating an efficiency effect. The fix is to relax the offset and let the coding-time coefficient be estimated from the data rather than imposed. We enter coding time in levels (rather than logs) so that zero coding-hour weeks participate naturally.

\subsection{PPML Specification}

Given the failure of the proportionality assumption, we estimate GHCP's association with PR output using Poisson Pseudo-Maximum Likelihood (PPML) with absorbed engineer and week fixed effects. Coding time and browser time (measures of development effort) enter in levels, so zero coding-hour weeks participate naturally and the model learns the effort-to-PRs relationship without forcing proportionality. Engineer FE ($\alpha_i$) ensure identification from within-person variation and week FE ($\delta_t$) absorb org-wide shocks. PPML is robust to distributional misspecification (overdispersion, zero-inflation, etc.), producing consistent estimates even when the data violate standard Poisson assumptions \cite{santos2006}. All standard errors are clustered at the manager level. This is more conservative than engineer-level clustering because it permits arbitrary within-team correlation in residuals, accommodating shared shocks such as release cadence, sprint pressure, and task allocation that a manager sets for the team. Engineer fixed effects already absorb time-invariant within-person correlation, so clustering at the manager level targets the residual cross-engineer dependence that engineer-level clustering would miss.

\subsection{Estimand: Efficiency, Not Total Effect}
\label{sec:estimand}

Conditioning on coding time and browser time is not only a confounding fix, it is an estimand choice. If GHCP reduces the time engineers need to complete a given task (the most natural expectation for a labor-saving tool), then coding time is partly a \emph{mediator} on the causal path from GHCP to PR output, not a pure confounder. Conditioning on a mediator blocks that indirect channel: any GHCP-induced time savings are absorbed by the controls rather than captured in the estimate. In the causal mediation literature, the resulting quantity corresponds to a \emph{controlled direct effect} \cite{acharya2016}: the effect of the treatment holding the mediator fixed at a given level.

The resulting estimand is an efficiency effect: the change in completed PRs per unit of measured development effort attributable to GHCP usage.

The total effect is the sum of (a) this efficiency channel and (b) the time-savings channel through which GHCP reduces the effort required to complete a task. Because our specification conditions on effort, channel (b) is absorbed. The practical implication is that our coefficients may underestimate the full impact of GHCP on PR throughput. The tool may both raise output per hour \emph{and} free up hours, but we measure only the former. If GHCP does indeed yield time savings, then our estimates understate the total productivity impact.

Further, we view this as the more actionable estimand. It answers "does GHCP make each coding hour more productive?" rather than conflating productivity gains with shifts in time allocation, and that question is directly actionable regardless of how engineers reallocate any time savings. Under the assumption that time freed by GHCP is not systematically reallocated to non-PR work, the efficiency effect is also a lower bound on the total PR throughput effect.

\subsection{Data Construction}
\label{sec:data}

\textbf{Pull requests.} PRs are attributed to the week they were created, not merged. This anchors the outcome to the period of active work, aligning it with contemporaneous effort and GHCP usage measures. A 28-day completion window ensures equal opportunity for PRs created in any study week to reach completed status. PRs still open beyond 28 days are excluded.

\textbf{GHCP telemetry.} Our GHCP usage measure aggregates telemetry across GitHub Copilot's IDE integrations. The GitHub Copilot CLI is not included as it was not generally available for most of the study window. For the Feb--Dec 2025 period this captures the dominant modes of GHCP usage in our population.

\textbf{Activity measures.} Coding time (time in any IDE), browser time (time in any web browser), and total computer activity (time across all applications) are weekly per-engineer aggregates from Microsoft's internal interactivity telemetry, which records active engagement time with applications on the engineer's workstation. Total computer activity is used only for the sample filter described next.

\textbf{Sample restrictions.} Two filters ensure measurement validity. At the \emph{engineer level}, we exclude individuals with either zero completed PRs or zero GHCP usage days across the full 43-week period. Zero PRs likely reflects a measurement gap (e.g., contributions to open source repositories), while zero GHCP days likely indicates use of an alternative coding assistant rather than unassisted development. This restriction is conservative: the PPML estimator would drop these engineers regardless, as they contribute no within-person variation. At the \emph{week level}, engineer-weeks with fewer than 10 hours of total computer activity are excluded as vacation or leave. This filter applies to overall activity, not coding specifically, so working weeks with zero coding hours are retained and enter the model with coding hours = 0.

\textbf{Population.} 16,223 individual contributor Software Engineers within Microsoft's Cloud+AI organization, observed over 43 weeks (Feb 23 -- Dec 21, 2025).

\subsection{Data Distribution}

\begin{figure}[H]
\begin{center}
\centerline{\includegraphics[width=\columnwidth]{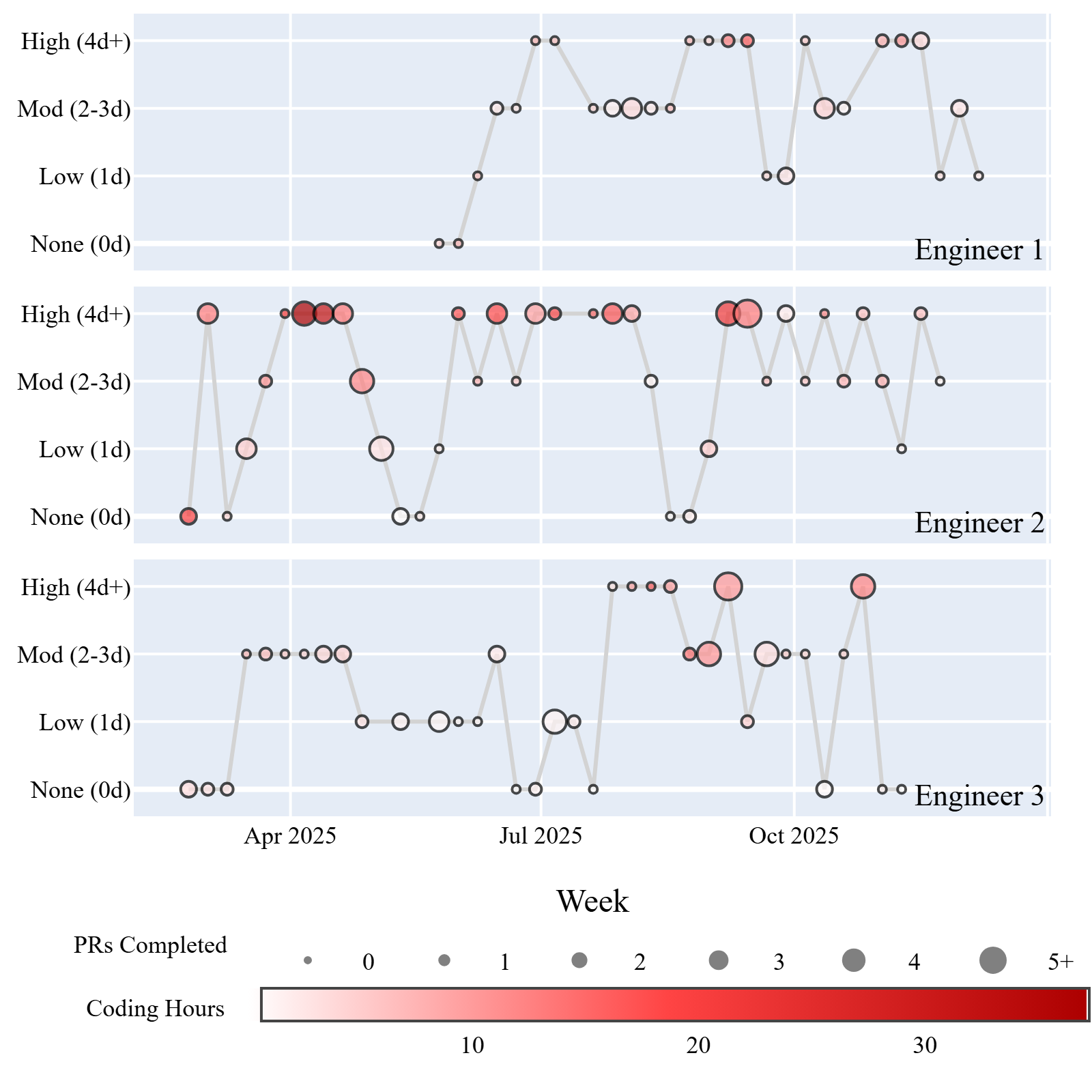}}
\caption{Weekly GHCP usage trajectories for three sampled engineers. Dot size = PRs completed; dot color = coding hours.}
\label{fig:trajectories}
\end{center}
\end{figure}

Figure~\ref{fig:trajectories} traces three randomly sampled engineers across all 43 weeks. Each point marks one engineer-week, visualizing GHCP usage, PRs completed, and hours spent coding. Engineers move substantially between usage levels from week to week (the within-engineer variation the model exploits), and PR-heavy weeks tend to cluster in the higher GHCP usage rows.

\newpage
\section{Results}
\label{sec:results}
\label{sec:depth}

We operationalize GHCP usage as \textbf{interaction depth}: total GHCP suggestions, prompts, and accepts per week. Splitting non-zero usage into terciles yields four levels:

\begin{itemize}
\item \textbf{Zero Usage:} 0 interactions (32.5\% of engineer-weeks)
\item \textbf{Low:} 1--41 interactions (22.6\%)
\item \textbf{Moderate:} 42--164 interactions (22.5\%)
\item \textbf{High:} 165+ interactions (22.4\%)
\end{itemize}

\noindent Interaction count measures actual GHCP engagement volume conditional on working, making it a treatment variable that is not mechanically tied to attendance (an alternative breadth-based operationalization, days per week with any GHCP usage, is examined as a robustness check in Section~\ref{sec:breadth}). The full panel of 413,732 engineer-weeks enters PPML estimation:

\begin{align*}
\ln E[\text{PRs}_{it}] &= \beta_1 \cdot \text{CodingTime}_{it} + \beta_2 \cdot \text{BrowserTime}_{it} \\
&\quad + \beta_3 \cdot \text{Low}_{it} + \beta_4 \cdot \text{Mod}_{it} + \beta_5 \cdot \text{High}_{it} \\
&\quad + \delta_t + \alpha_i
\end{align*}

\textbf{Result:} Engineers complete 40.5\% more pull requests in their highest GHCP usage weeks compared to their zero-usage weeks, at similar levels of coding and browser time (Table~\ref{tab:depth}). The gradient is monotonic (Low +21\%, Moderate +39\%, High +41\%) but flattens from Moderate to High. This saturation pattern is consistent with diminishing returns, whereas a perfectly linear gradient would be harder to distinguish from mechanical confounding.

\begin{figure}[H]
\begin{center}
\centerline{\includegraphics[width=\columnwidth]{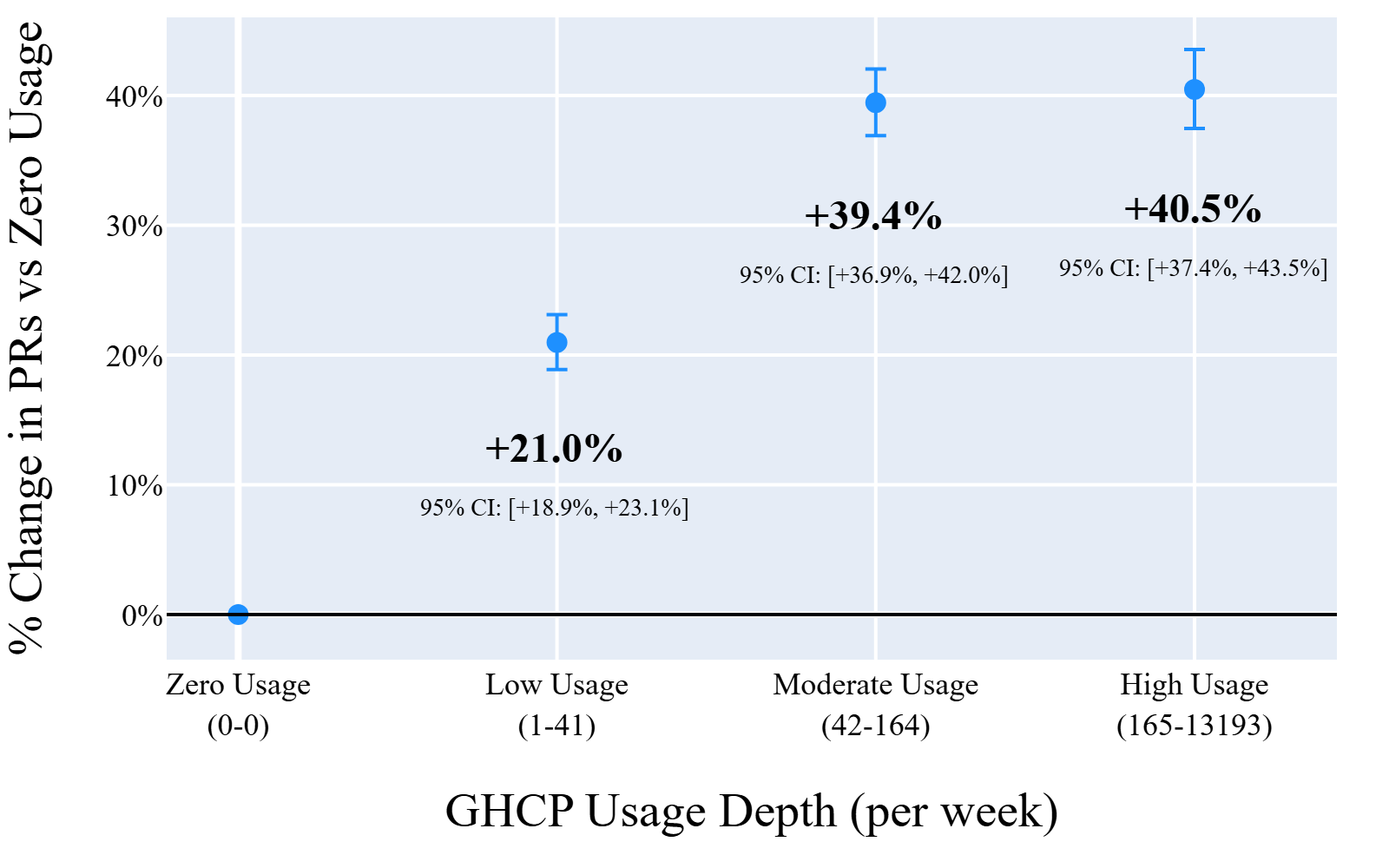}}
\caption{Usage depth dose-response.}
\label{fig:depth}
\end{center}
\end{figure}

\newpage

\begin{table}[H]
\centering
\footnotesize
\caption{Depth-based dose-response, PPML with two-way fixed effects.}
\label{tab:depth}
\begin{threeparttable}
\renewcommand{\arraystretch}{1.05}
\setlength{\tabcolsep}{6pt}
\begin{tabular}{lcc}
\toprule
 & \multicolumn{2}{c}{Completed PRs} \\
\cmidrule(lr){2-3}
 & Log scale & \% change \\
 & $\beta$ (SE) & $\exp(\beta)-1$ (SE, pp) \\
\midrule
Coding time (hours) & 0.021 (0.001) & 2.2\% (0.1) \\
Browser time (hours) & 0.031 (0.001) & 3.1\% (0.1) \\
Low GHCP usage & 0.190 (0.009) & 21.0\% (1.1) \\
Moderate GHCP usage & 0.332 (0.009) & 39.4\% (1.3) \\
High GHCP usage & 0.340 (0.011) & 40.5\% (1.6) \\
\midrule
Engineer FE & \multicolumn{2}{c}{\checkmark} \\
Week FE     & \multicolumn{2}{c}{\checkmark} \\
\midrule
Observations & \multicolumn{2}{c}{413,732} \\
Engineers    & \multicolumn{2}{c}{16,223} \\
Weeks        & \multicolumn{2}{c}{43} \\
\bottomrule
\end{tabular}
\begin{tablenotes}
\footnotesize
\item PPML (\texttt{fepois}) with engineer and week fixed effects. Standard errors clustered at the manager level. Coefficients are on the log scale; the percentage change in expected PR count equals $\exp(\beta) - 1$, with delta-method standard errors in percentage points.
\end{tablenotes}
\end{threeparttable}

\end{table}


\section{Robustness \& Falsification}
\label{sec:robustness}

The depth-based dose-response in Section~\ref{sec:depth} is the primary finding. Its causal interpretation rests on Assumption~\ref{assump:identification}, which cannot be tested directly. This section probes the assumption's plausibility through seven falsification tests, each targeting a distinct violation that could produce the observed gradient without a true productivity effect. Each subsection follows the same structure: identify the threat, design a sharp test, and report the result.

\newpage
\subsection{Placebo Treatment}
\label{sec:placebo-treatment}

\textbf{Threat:} Generic engagement with AI tools, not GHCP specifically. In high-engagement weeks, engineers may use more AI across many surfaces and also complete more PRs for reasons unrelated to coding assistance. If this latent ``AI-engaged week'' state drives the result, then non-coding Copilot usage should show a similar dose-response with PR output.

\textbf{Test:} Keep the PR outcome and full headline specification fixed, but replace GHCP usage with non-coding M365 Copilot usage in Word, Excel, PowerPoint, Teams, and Outlook. These surfaces capture general AI engagement but have little direct pathway to completed PRs. We operationalize the placebo treatment as breadth (days per week with any non-coding Copilot usage) because depth-equivalent interaction counts are not available across all M365 surfaces. This is conservative, since breadth is the more attendance-correlated measure and therefore more likely to spuriously track PR output if any engagement signal exists.

\begin{align*}
\ln E[\text{PRs}_{it}] &= \beta_1 \cdot \text{CodingTime}_{it} + \beta_2 \cdot \text{BrowserTime}_{it} \\
&\quad + \beta_3 \cdot \text{Low}^{\text{M365}}_{it} + \beta_4 \cdot \text{Mod}^{\text{M365}}_{it} \\
&\quad + \beta_5 \cdot \text{High}^{\text{M365}}_{it} + \delta_t + \alpha_i
\end{align*}

\textbf{Result:} The non-coding M365 Copilot gradient is essentially flat across all usage levels. Point estimates oscillate narrowly around zero (Low and Moderate marginally negative, High marginally positive at roughly +3\%), with confidence intervals that span or sit adjacent to zero. The GHCP depth gradient under the headline specification, by contrast, rises monotonically to +40\%. Within-engineer week-level non-coding AI engagement does not predict PR output---coding-specific assistance does.

\begin{figure}[H]
\begin{center}
\centerline{\includegraphics[width=\columnwidth]{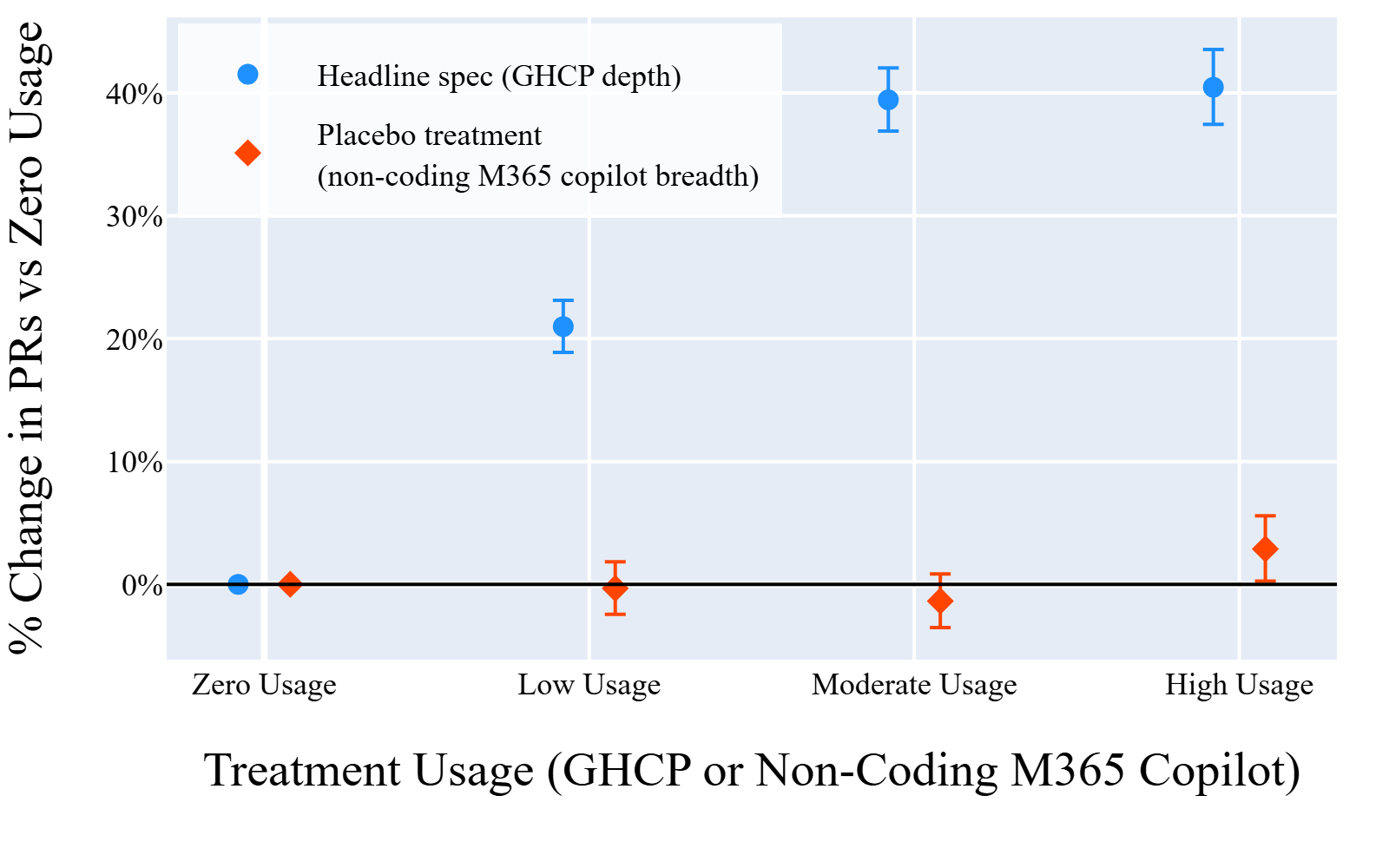}}
\caption{Placebo treatment: non-coding M365 Copilot breadth vs.\ GHCP interaction depth as the treatment variable.}
\label{fig:placebo-treatment}
\end{center}
\end{figure}

\newpage
\subsection{Placebo Outcome}
\label{sec:placebo-outcome}

\textbf{Threat:} The base controls absorb individual effort variation (coding and browser time) and org-wide shocks (week fixed effects), but neither addresses team-specific pressures (a sprint deadline or release push that hits a specific team). These team-level shocks
could simultaneously increase both an individual's GHCP usage and their PR output, inflating the gradient for reasons unrelated to GHCP itself.

\textbf{Test:} Keep the real treatment (individual GHCP usage levels) but replace the outcome with the leave-one-out team PR count (PRs completed by the individual's teammates that week). If the gradient is driven by shared team shocks, individual GHCP usage should predict teammates' PRs at comparable magnitudes.

\begin{align*}
\ln E[\text{TeamPRs}_{it}] &= \beta_1 \cdot \text{CodingTime}_{it} + \beta_2 \cdot \text{BrowserTime}_{it} \\
&\quad + \beta_3 \cdot \text{Low}_{it} + \beta_4 \cdot \text{Mod}_{it} + \beta_5 \cdot \text{High}_{it} \\
&\quad + \delta_t + \alpha_i
\end{align*}

\textbf{Result:} The placebo gradient is negligible. GHCP usage coefficients on teammates' PRs are on the order of 1--2\%, compared to 20--40\% for the individual's own PRs. This weighs against shared team-level shocks as the confound.

\begin{figure}[H]
\begin{center}
\centerline{\includegraphics[width=\columnwidth]{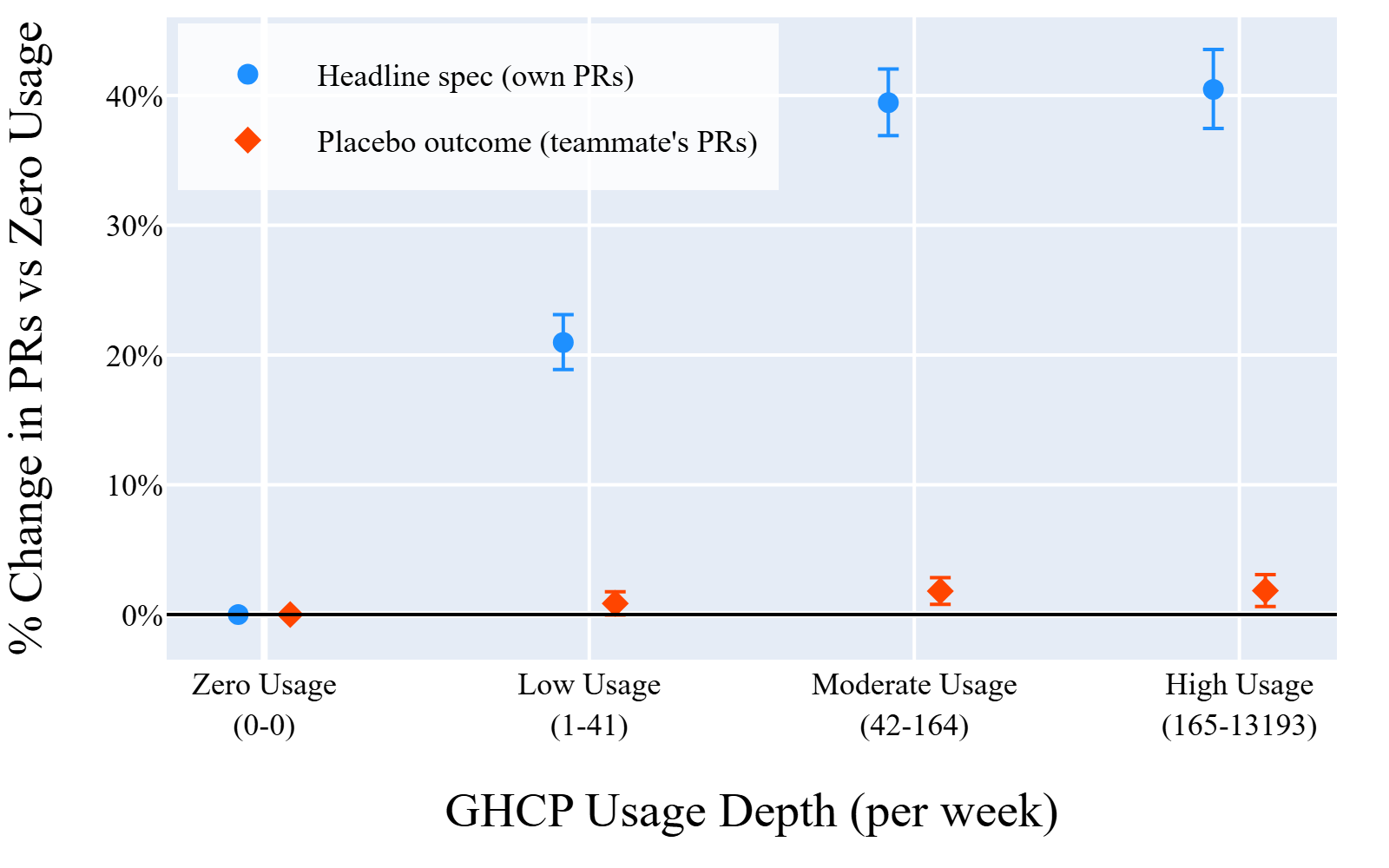}}
\caption{Placebo outcome: own PRs vs.\ teammates' PRs by GHCP usage depth.}
\label{fig:placebo-outcome}
\end{center}
\end{figure}

\newpage
\subsection{Task-Mix Shift}
\label{sec:task-mix}

\textbf{Threat:} Reallocation, not productivity. Active coding time is a finite weekly budget split across several activities, most prominently authoring one's own PRs and reviewing teammates' PRs. If high-GHCP weeks are simply weeks in which an engineer happens to spend more of that fixed budget authoring rather than reviewing, the headline gradient could reflect a within-week reshuffling of \emph{what kind} of work gets done rather than an increase in \emph{how much} work gets done. Under this story, GHCP would simply be coinciding with weeks dominated by creation tasks rather than making engineers more productive.

\textbf{Test:} The reallocation story makes a sharp directional prediction. If a week's coding hours are a fixed pie split between authoring and reviewing, then any increase in PRs authored at high GHCP usage must be offset by a \emph{decrease} in PRs reviewed at the same usage level. We test this by re-estimating the headline specification with PR reviews given as the outcome, holding all controls and the treatment definition fixed.

\begin{align*}
\ln E[\text{ReviewsGiven}_{it}] &= \beta_1 \cdot \text{CodingTime}_{it} + \beta_2 \cdot \text{BrowserTime}_{it} \\
&\quad + \beta_3 \cdot \text{Low}_{it} + \beta_4 \cdot \text{Mod}_{it} + \beta_5 \cdot \text{High}_{it} \\
&\quad + \delta_t + \alpha_i
\end{align*}

\begin{figure}[H]
\begin{center}
\centerline{\includegraphics[width=\columnwidth]{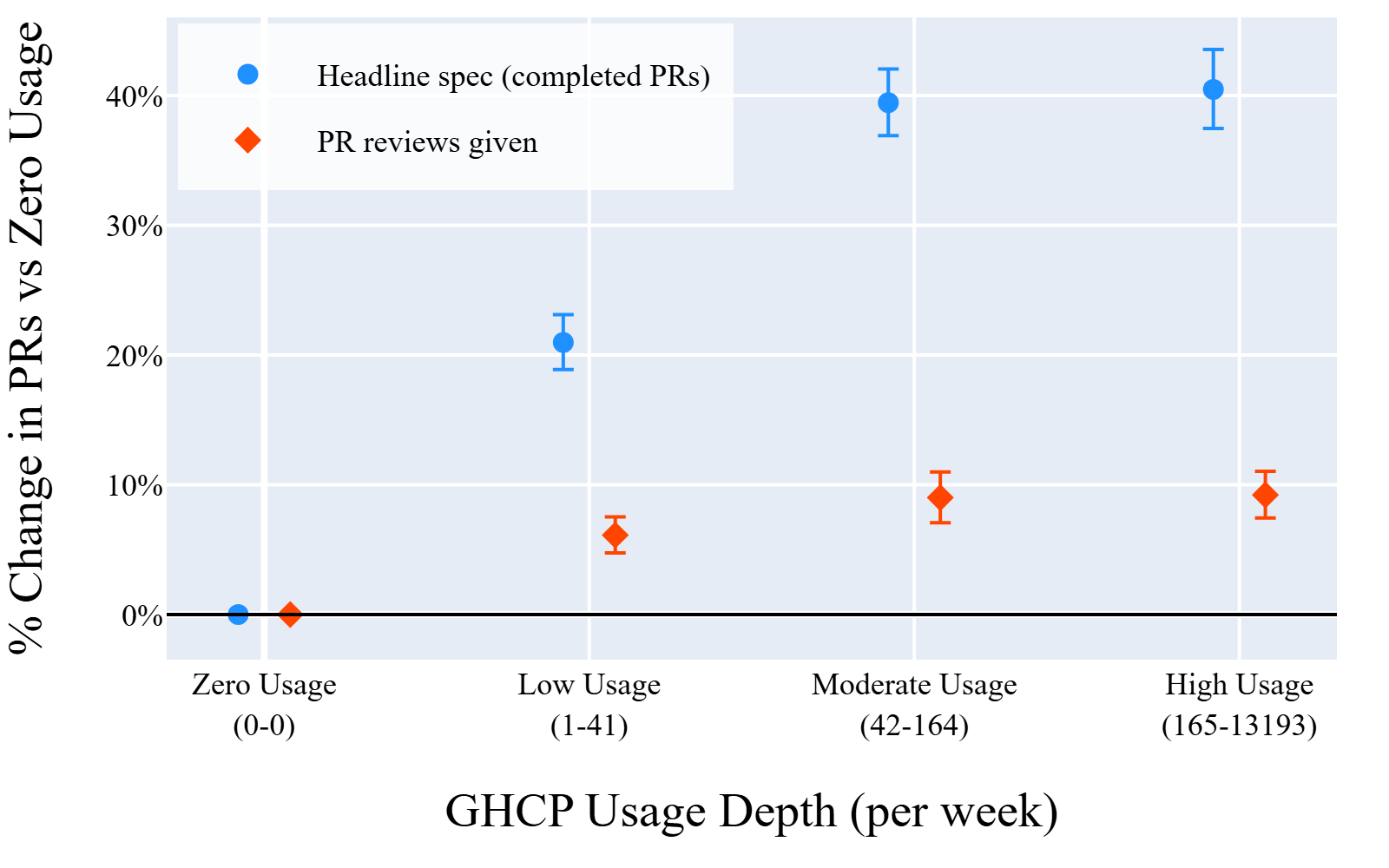}}
\caption{Task-mix test: GHCP gradient on PRs authored vs.\ reviews given, holding the specification fixed.}
\label{fig:task-mix}
\end{center}
\end{figure}

\textbf{Result:} The reviews-given gradient is monotonically positive across the GHCP usage levels, mirroring the authored-PRs gradient rather than offsetting it (Figure~\ref{fig:task-mix}). Both throughput measures rise together at higher GHCP usage, which is the opposite of the substitution pattern this version of the task-mix story requires. A within-week reallocation between authoring and reviewing is therefore unlikely to account for the headline gradient, though this test does not rule out other task-mix shifts (e.g., toward qualitatively easier work) that would not produce an authoring-reviewing tradeoff.

\newpage
\subsection{Timing Tests}
\label{sec:lead-test}

\textbf{Threat:} Multi-week productive states drive both GHCP usage and PR output across consecutive weeks, contaminating the contemporaneous estimate.

\textbf{Test:} Fit two models, each adding one displaced GHCP usage measure alongside the current-week levels. A lag model includes the prior week's usage ($t-1$) and a lead model includes the following week's usage ($t+1$). If the association is truly contemporaneous, neither displaced measure should have meaningful positive predictive power once we condition on this week's usage.

\begin{align*}
\ln E[\text{PRs}_{it}] &= \beta_1 \cdot \text{CodingTime}_{it} + \beta_2 \cdot \text{BrowserTime}_{it} \\
&\quad + \beta_3 \cdot \text{Low}_{it} + \beta_4 \cdot \text{Mod}_{it} + \beta_5 \cdot \text{High}_{it} \\
&\quad + \beta_6 \cdot \text{Low}_{i,t-1} + \beta_7 \cdot \text{Mod}_{i,t-1} \\
&\quad + \beta_8 \cdot \text{High}_{i,t-1} + \delta_t + \alpha_i
\end{align*}

\textbf{Result:} Neither displaced measure is positively predictive. The lag coefficients are near zero, meaning last week's GHCP usage adds essentially no information. The lead coefficients are slightly negative, consistent with mean reversion. A positive association does not bleed across week boundaries, which is not the pattern a persistent unobserved state explanation would predict.

\begin{figure}[H]
\begin{center}
\centerline{\includegraphics[width=\columnwidth]{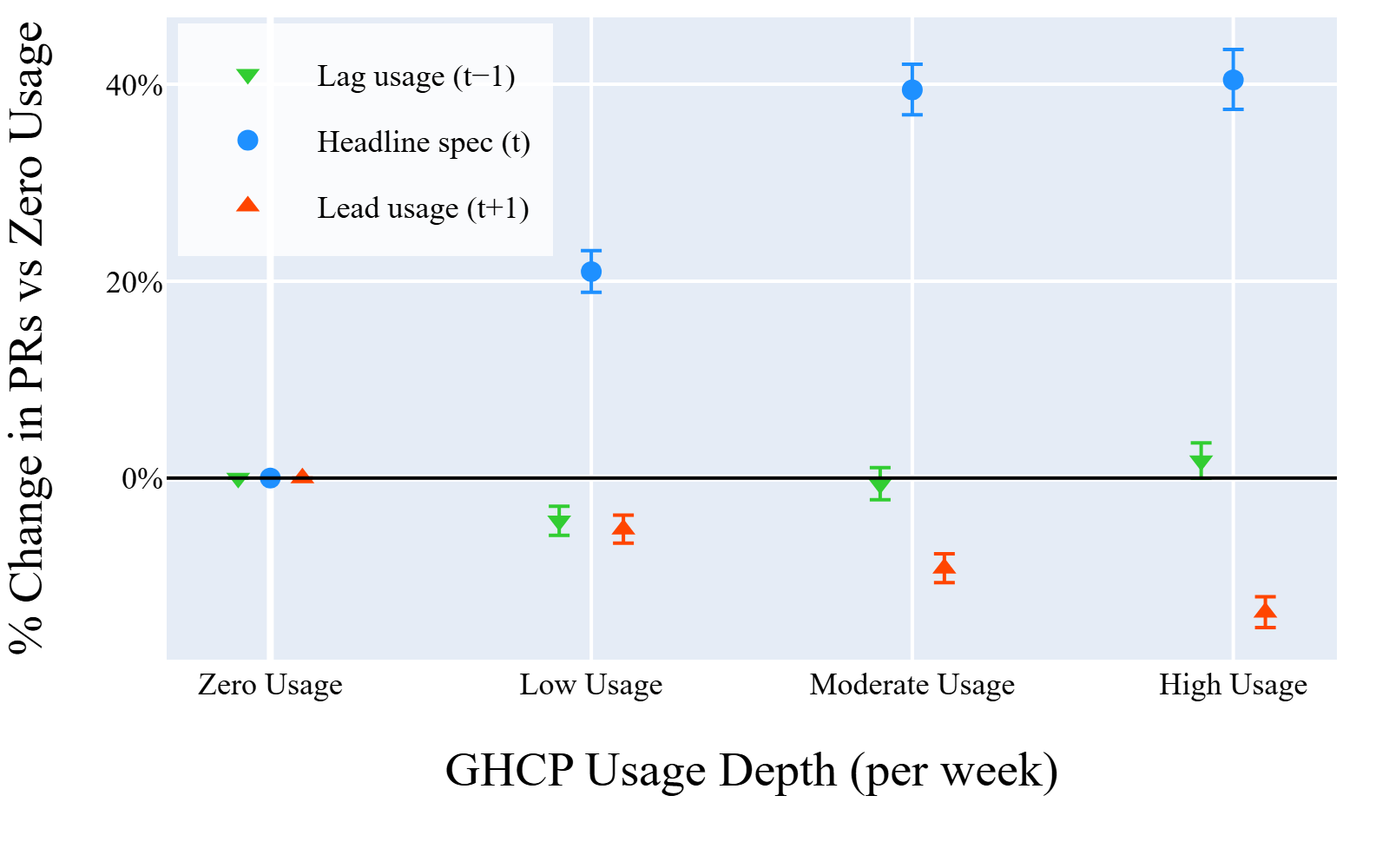}}
\caption{Timing tests: current ($t$), lagged ($t{-}1$), and leading ($t{+}1$) GHCP usage coefficients.}
\label{fig:timing}
\end{center}
\end{figure}

\newpage
\subsection{PR Size Decomposition}
\label{sec:pr-size-mix}

\textbf{Threat:} Slicing, not productivity. The headline gradient counts PRs without regard to their size. If engineers in high-GHCP weeks simply chop the same body of work into more, smaller PRs, the gradient would reflect a shift in unit size rather than an increase in substantive engineering throughput. Under this story the throughput gain should concentrate in small PRs and weaken or disappear for larger ones.

\textbf{Test:} Decompose the Azure DevOps (ADO) PR count into three sub-counts by files touched, where ``files touched'' is the sum of files added, changed, and deleted in the PR (small: 1 file; medium: 2--6 files; large: 7+ files). Fit the depth specification separately to each of these three count outcomes on the identical engineer-week panel with identical controls:

\begin{align*}
\ln E[\text{PRs}^{s}_{it}] &= \beta_1 \cdot \text{CodingTime}_{it} + \beta_2 \cdot \text{BrowserTime}_{it} \\
&\quad + \beta_3 \cdot \text{Low}_{it} + \beta_4 \cdot \text{Mod}_{it} + \beta_5 \cdot \text{High}_{it} \\
&\quad + \delta_t + \alpha_i, \quad s \in \{\text{small, medium, large}\}
\end{align*}

\textbf{Result:} The dose-response steepens with PR size. The small- and medium-PR gradients both rise from Low to Moderate usage and then plateau or attenuate at High usage, while the large-PR gradient is monotonically increasing and the steepest of the three (+11\% / +47\% / +70\%). If GHCP throughput gains were driven by engineers slicing the same work into more, smaller PRs, the small-PR series would dominate; instead, the effect is concentrated in PRs touching seven or more files, which are less easily dismissed as trivial slicing.

\begin{figure}[H]
\begin{center}
\centerline{\includegraphics[width=\columnwidth]{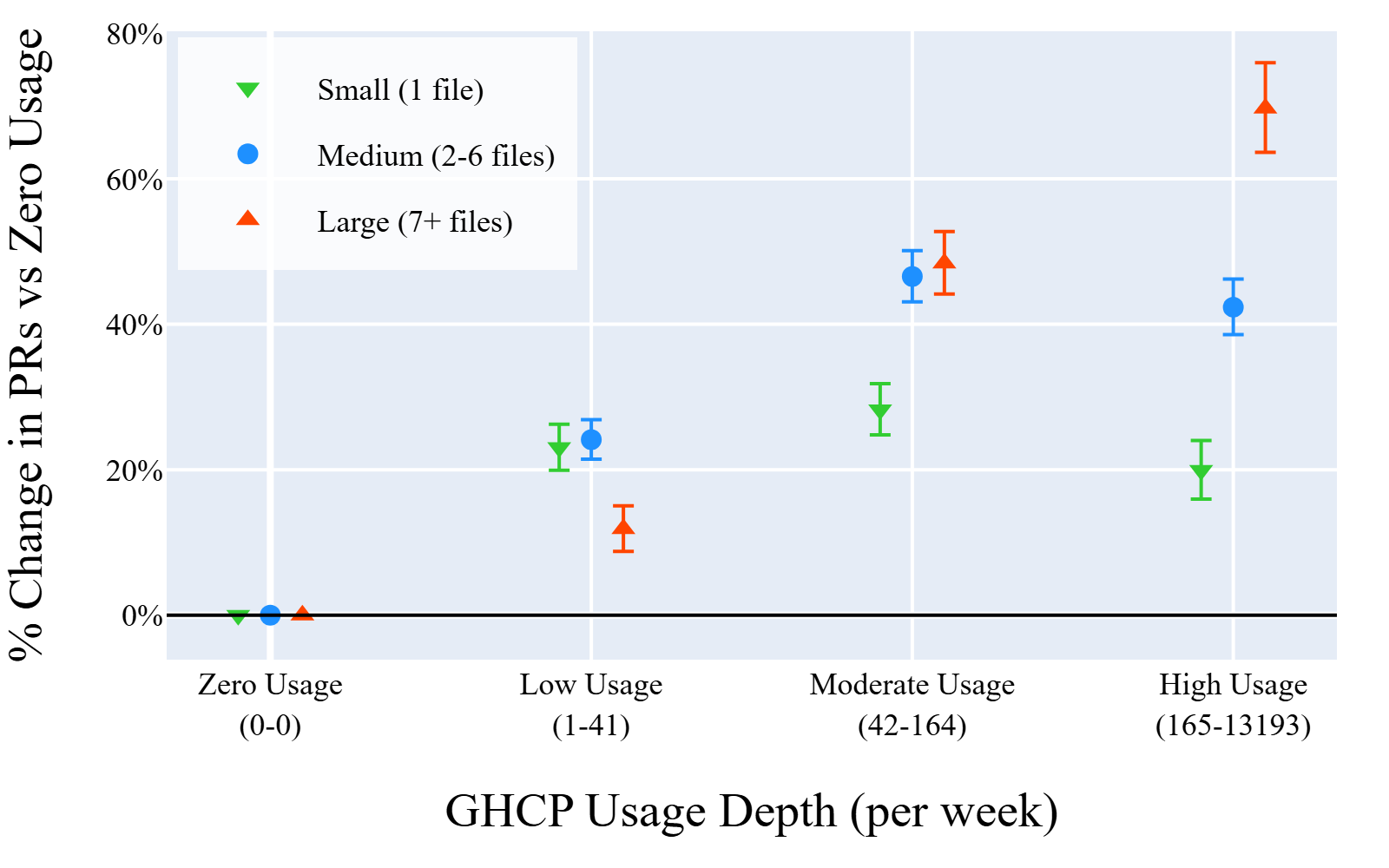}}
\caption{PR size decomposition: dose-response by files touched (ADO subsample).}
\label{fig:pr-size}
\end{center}
\end{figure}

\newpage
\subsection{PR-Type Decomposition}
\label{sec:pr-type}

\textbf{Threat:} Easy work, not productivity. File count is an imperfect proxy for task difficulty: a PR touching several configuration or documentation files is typically easier than one touching a single source file. If high-GHCP weeks coincide with more config/doc work, the headline gradient could reflect a shift toward easier tasks rather than substantive throughput.

\textbf{Test:} Decompose the ADO PR count into two parallel sub-counts by file extension composition. A PR is classified as \emph{config/\allowbreak documentation-only} if every file it touches has an extension in a curated config/doc set (e.g., \texttt{.md}, \texttt{.json}, \texttt{.yaml}, \texttt{.yml}, \texttt{.txt}, \texttt{.xml}). All remaining PRs, including those touching at least one source code file, form the \emph{other PRs} outcome. We refit the depth specification separately to each outcome on the identical engineer-week panel with identical controls:

\begin{align*}
\ln E[\text{PRs}^{\tau}_{it}] &= \beta_1 \cdot \text{CodingTime}_{it} + \beta_2 \cdot \text{BrowserTime}_{it} \\
&\quad + \beta_3 \cdot \text{Low}_{it} + \beta_4 \cdot \text{Mod}_{it} + \beta_5 \cdot \text{High}_{it} \\
&\quad + \delta_t + \alpha_i, \quad \tau \in \{\text{config/doc, other}\}
\end{align*}

\noindent If the threat holds, the gradient should be concentrated in config/\allowbreak documentation-only PRs, with the ``other PRs'' gradient flat or substantially attenuated.

\textbf{Result:} The ``other PRs'' gradient is monotonic (+21\% / +51\% / +58\%) and actually steeper at moderate and high usage than the pooled headline gradient. Because this series excludes config/\allowbreak documentation-only PRs by construction, the headline gradient is unlikely to be a reallocation toward easier task types; if anything, removing those PRs sharpens the dose-response on the remaining slice, the opposite of what the threat predicts.

\begin{figure}[H]
\begin{center}
\centerline{\includegraphics[width=\columnwidth]{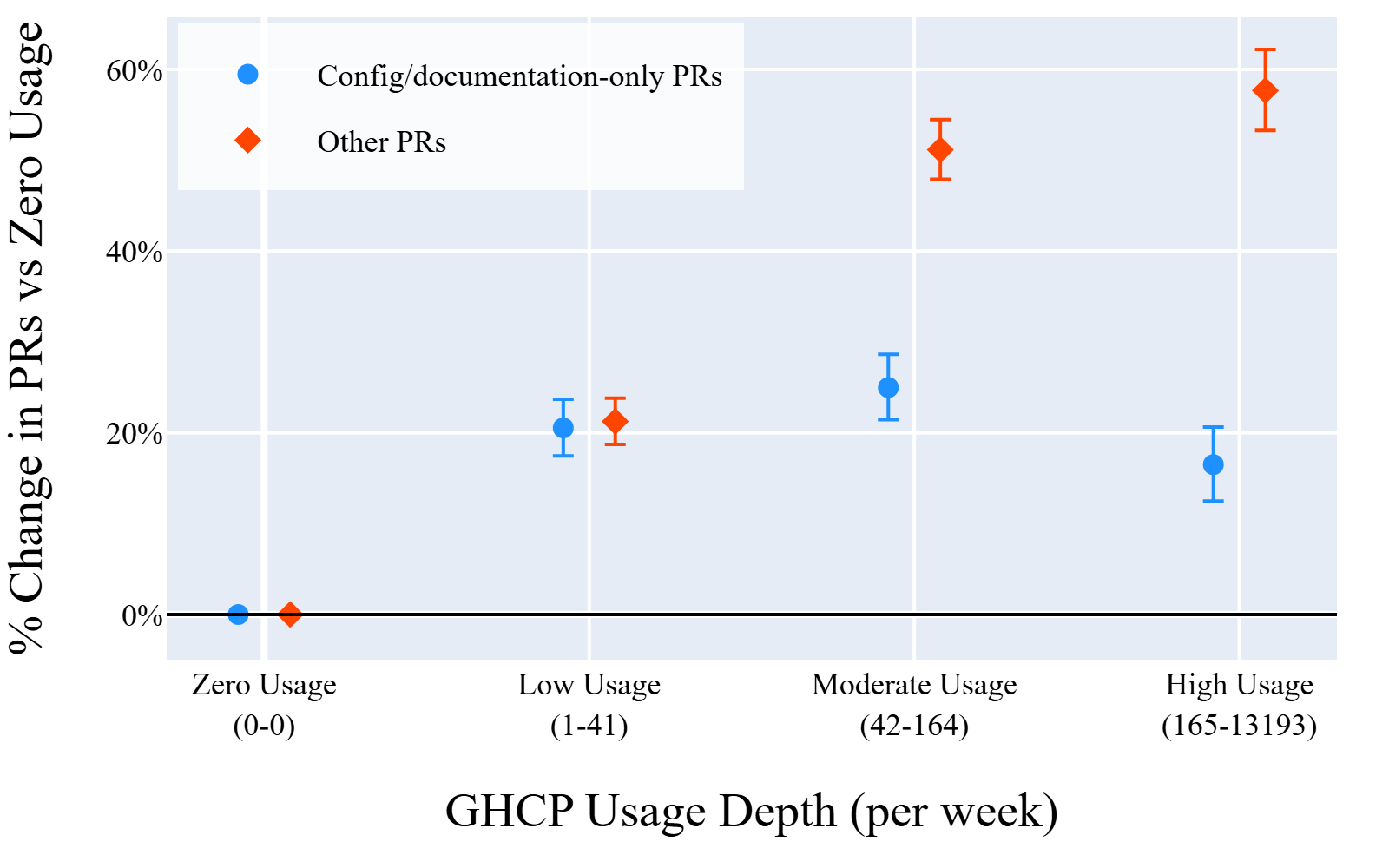}}
\caption{PR-type decomposition: dose-response for config/\allowbreak documentation-only PRs vs.\ all other PRs (ADO subsample).}
\label{fig:pr-type}
\end{center}
\end{figure}

\newpage
\subsection{Alternative Operationalization}
\label{sec:breadth}

\textbf{Threat:} The depth-based bucketing (terciles of non-zero weekly interactions) is one of many defensible ways to operationalize GHCP usage. The headline gradient could in principle be sensitive to that particular choice.

\textbf{Test:} Refit the headline specification with breadth in place of depth, where breadth is days per week with at least one GHCP interaction, bucketed into None (0d), Low (1d), Moderate (2--3d), and High (4d+). Breadth and depth are correlated but capture different aspects of usage: breadth is largely a proxy for days spent coding, while depth measures engagement volume conditional on working. If the depth result is an artifact of bucketing, the breadth gradient should look qualitatively different.

\begin{align*}
\ln E[\text{PRs}_{it}] &= \beta_1 \cdot \text{CodingTime}_{it} + \beta_2 \cdot \text{BrowserTime}_{it} \\
&\quad + \beta_3 \cdot \text{Low}^{\text{br}}_{it} + \beta_4 \cdot \text{Mod}^{\text{br}}_{it} + \beta_5 \cdot \text{High}^{\text{br}}_{it} \\
&\quad + \delta_t + \alpha_i
\end{align*}

\textbf{Result:} The breadth gradient is also monotonically increasing, with high-breadth weeks associated with a +54\% increase in PRs relative to zero-breadth weeks. The two operationalizations converge on the same qualitative story, suggesting the headline finding is not sensitive to this particular alternative operationalization.

\begin{figure}[H]
\begin{center}
\centerline{\includegraphics[width=\columnwidth]{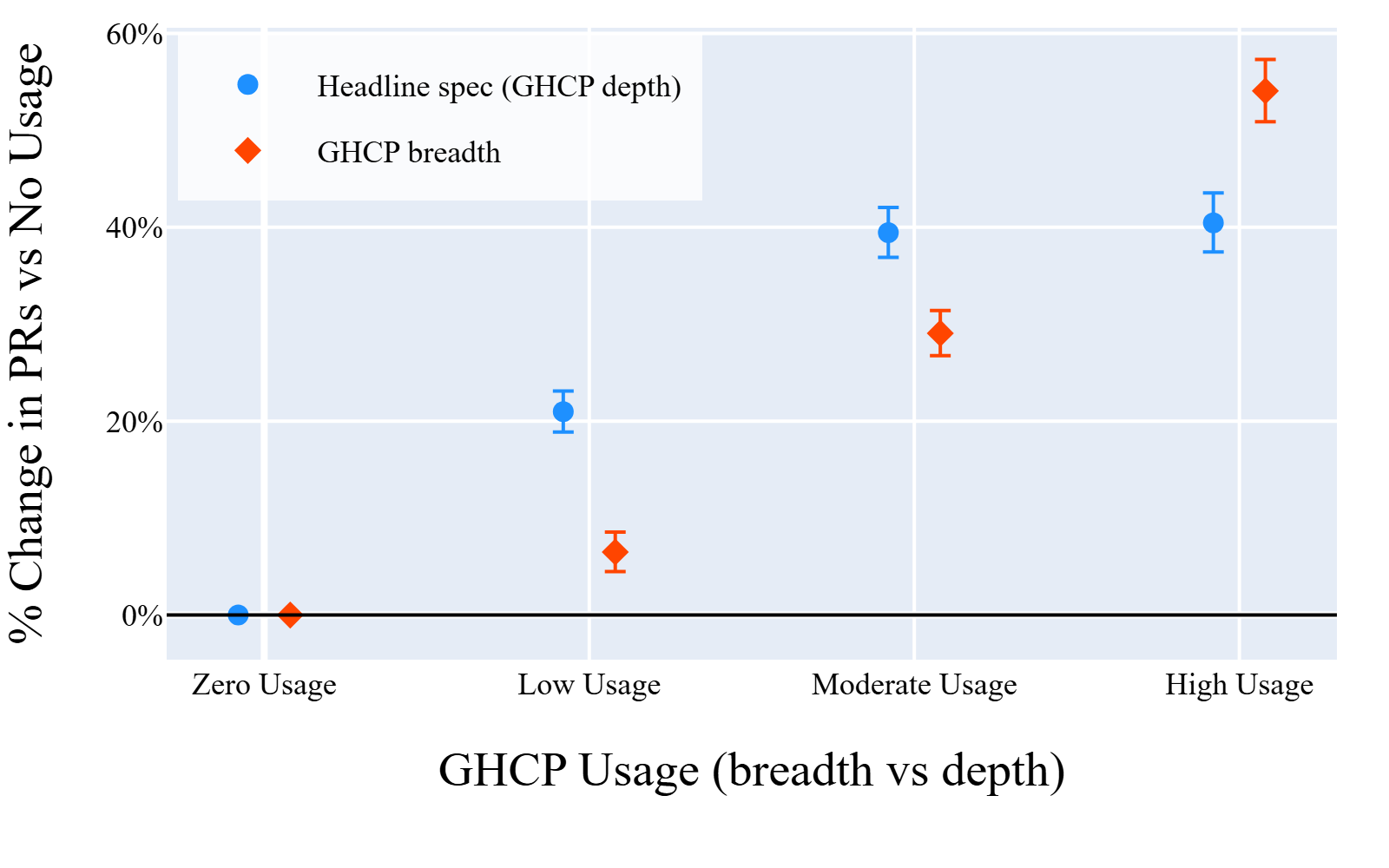}}
\caption{Alternative treatment operationalization: usage breadth dose-response.}
\label{fig:breadth}
\end{center}
\end{figure}

\newpage
\subsection{Robustness Summary}

No single test is decisive, but together they narrow the space of plausible alternative explanations. The two placebo tests address complementary threats: Section~\ref{sec:placebo-outcome} weighs against team-level shocks (the real treatment does not predict a placebo outcome), and Section~\ref{sec:placebo-treatment} weighs against generic within-engineer AI engagement (a non-coding AI placebo treatment produces no gradient under the same specification). The task-mix test (Section~\ref{sec:task-mix}) is hard to reconcile with a within-week reallocation between authoring and reviewing PRs: both throughput measures rise together with GHCP usage, the opposite of the substitution pattern that story requires. The two decomposition tests sharpen the result rather than merely sustain it. Decomposing by PR size (Section~\ref{sec:pr-size-mix}) shows the gradient is steepest for large (7+ file) PRs, the opposite of what a ``slicing into smaller PRs'' story predicts; decomposing by PR type (Section~\ref{sec:pr-type}) shows that excluding configuration- and documentation-only PRs strengthens the dose-response on the remaining slice, the opposite of what an ``easier task mix'' story predicts. The timing tests (Section~\ref{sec:lead-test}) indicate the association is contemporaneous rather than bleeding across week boundaries. The alternative operationalization (Section~\ref{sec:breadth}) suggests the gradient is unlikely to be an artifact of how depth is bucketed: a breadth-based measure produces the same monotonic pattern.

\section{Discussion}
\label{sec:discussion}

\subsection{Interpretation}

The depth-based model (Section~\ref{sec:depth}) supports a clear story: engineers complete more PRs in weeks when they use GHCP more intensively, after absorbing all time-invariant individual differences and controlling for coding time. The robustness battery narrows the space of plausible confounds substantially: the association is unlikely to be an artifact of within-week reallocation between authoring and reviewing PRs (Section~\ref{sec:task-mix}), team-level shocks (Section~\ref{sec:placebo-outcome}), generic AI engagement (Section~\ref{sec:placebo-treatment}), smaller PR scope (Section~\ref{sec:pr-size-mix}), a shift toward configuration- or documentation-only work (Section~\ref{sec:pr-type}), or how the treatment is bucketed (Section~\ref{sec:breadth}), and the timing tests (Section~\ref{sec:lead-test}) indicate the association is contemporaneous rather than driven by cross-week contamination.

\newpage
\subsection{Limitations}

\textbf{Remaining threat.} The remaining threats are unobservable, week-to-week shifts that influence both GHCP use and PR throughput simultaneously: qualitatively easier tasks (beyond what file-count metrics capture) or bursts of intrinsic motivation (beyond what coding time captures). These are precisely the violations that Assumption~\ref{assump:identification} rules out. The robustness battery probes their plausibility but cannot verify the assumption directly.

\textbf{PR count as outcome.} Developer productivity is inherently multi-dimensional \cite{forsgren2021}. Our outcome is completed PRs, which captures only the activity and efficiency dimensions, leaving satisfaction, performance, and communication unmeasured. Section~\ref{sec:pr-size-mix} partially addresses PR complexity by decomposing the outcome by files touched, and Section~\ref{sec:pr-type} further decomposes it by file type. The stability of the gradient across these decompositions suggests the throughput increase is not purely a task-decomposition artifact, but these tests are limited to ADO-hosted repositories and cannot fully capture PR complexity.

\textbf{Contemporaneous measurement.} Treatment and outcome are measured in the same week. Coding time and browser time control for the shared ``how much development happened'' signal, but a strictly causal interpretation requires that, after accounting for how much an engineer worked and all individual and time-period differences, there is nothing left over that simultaneously pushes both GHCP usage and PR output up or down in the same week.

\textbf{PR-development lead time.} A PR created in week $t$ may reflect work begun in $t{-}1$ or earlier, misaligning the outcome with the week of GHCP usage that assisted it. Two aggregate signatures weigh against this being a dominant artifact. First, prior-week GHCP usage adds essentially no predictive power once current-week usage is conditioned on (Section~\ref{sec:lead-test}), meaning the spillover that multi-week development would generate is not detectable. Second, the threat predicts the most attenuated gradient for long-running PRs, yet the dose-response is \emph{steepest} for large (7+ file) PRs (Section~\ref{sec:pr-size-mix}), which on average involve more work. These tests do not resolve lead time at the individual PR level, but the aggregate signatures the threat would produce are absent.

\textbf{Efficiency estimand, not total effect.} As discussed in Section~\ref{sec:estimand}, conditioning on coding time and browser time defines the estimand as an efficiency effect. If GHCP also reduces the time engineers need to complete a given workload, that time-savings channel is absorbed by the controls rather than captured in the estimate. To the extent that GHCP does yield such time savings, the total productivity effect would exceed the efficiency effect reported here.

\newpage
\subsection{Conclusion}

Does GitHub Copilot make engineers more productive, or are GHCP-heavy weeks systematically different from GHCP-light weeks for reasons unrelated to the tool? Engineer fixed effects address the easier version of this question---the engineers who use GHCP more are not just different people---by comparing each engineer against themselves. The harder version concerns within-engineer, time-varying confounding: whether busy weeks, crunch periods, or task-mix shifts jointly drive both GHCP usage and PR output. That is the question our effort controls, week fixed effects, and falsification battery are designed to address, and only a randomized experiment could close it definitively. We have argued that such an experiment is infeasible for measuring GHCP's overall effect. What we can do is substantially narrow the space of plausible explanations. Using within-engineer panel data with two-way fixed effects, we find that the same engineer completes roughly 40\% more PRs in their highest GHCP-usage weeks compared to their zero-usage weeks, holding time spent coding constant. The gradient is monotonic and exhibits diminishing returns at high usage levels, a pattern more consistent with a genuine productivity effect than with mechanical confounding. A seven-test robustness battery substantially narrows the space of common alternative explanations, including the demonstration that an alternative breadth-based operationalization of GHCP usage yields a similar gradient. The causal interpretation of the estimates requires Assumption~\ref{assump:identification}; the battery narrows the set of violations that could plausibly explain the gradient, but the assumption itself is not directly testable.


\bibliographystyle{ACM-Reference-Format}
\bibliography{references}

\end{document}